\documentclass[aps,pra,twocolumn,superscriptaddress,longbibliography]{revtex4-1}

\usepackage{graphicx}
\usepackage{amsmath}
\usepackage{amssymb}
\usepackage{dsfont}
\usepackage{xspace}
\usepackage{color}
\usepackage[usenames,dvipsnames]{xcolor}
\usepackage[colorlinks=true,linkcolor=blue,urlcolor=blue,citecolor=blue]{hyperref}
\usepackage[normalem]{ulem}
\usepackage{times}

\DeclareMathOperator{\im}{\mbox{Im}}

\newcommand{\nag}{{\phantom{\dagger}}}

\newcommand{\eqw}[1]{(\ref{#1})}
\newcommand{\eq}[1]{Eq.\thinspace{}(\ref{#1})}

\newcommand{\tab}[1]{Tab.\thinspace{}\ref{#1}}

\newcommand{\fig}[1]{Fig.\thinspace{}\ref{#1}}

\newcommand{\fc}[1]{({#1})}
\newcommand{\figc}[2]{Fig.\thinspace{}\ref{#1}\thinspace{}\fc{#2}}

\def\ket#1{\mathinner{|{#1}\rangle}}

\newcommand{\KC}{K${}_3$C${}_{60}$ }

\newcommand{\eqfigscl}[2]{\vcenter{\hbox{\includegraphics[scale=#1]{#2}}}}

\begin{document}

\title{Dynamical Cooper pairing in non-equilibrium electron-phonon systems}

\author{Michael Knap}%
\affiliation{Department of Physics, Walter Schottky Institute, and Institute for Advanced Study, Technical University of Munich, 85748 Garching, Germany}%

\author{Mehrtash Babadi}%
\affiliation{Institute for Quantum Information and Matter, Caltech, Pasadena, CA 91125, USA}%

\author{Gil Refael}%
\affiliation{Institute for Quantum Information and Matter, Caltech, Pasadena, CA 91125, USA}%

\author{Ivar Martin}%
\affiliation{Materials Science Division, Argonne National Laboratory, Argonne, IL 60439, USA}%

\author{Eugene Demler}%
\affiliation{Department of Physics, Harvard University, Cambridge MA 02138, USA}%

\date{\today}

\begin{abstract}
We analyze Cooper pairing instabilities in strongly driven electron-phonon systems. The light-induced non-equilibrium state of phonons results in a simultaneous increase of the superconducting coupling constant and the electron scattering. We demonstrate that the competition between these effects leads to an enhanced superconducting transition temperature in a broad range of parameters. Our results may explain the observed transient enhancement of superconductivity in several classes of materials upon irradiation with high intensity pulses of terahertz light, and may pave new ways for engineering high-temperature light-induced superconducting states.
\end{abstract}


\maketitle

The application of a strong electromagnetic drive has emerged as a powerful new way to manipulate material properties~\cite{orenstein_ultrafast_2012, zhang_dynamics_2014}. Long-range charge density wave order has been melted by light~\cite{schmitt_transient_2008, yusupov_coherent_2010, hellmann_ultrafast_2010, rohwer_collapse_2011}, insulators have been destroyed~\cite{rini_control_2007, hilton_enhanced_2007, liu_terahertz-field-induced_2012-1}, and the breaking of superconducting pairs has been observed~\cite{demsar_pair-breaking_2003, cortes_momentum-resolved_2011, graf_nodal_2011, smallwood_tracking_2012,matsunaga_higgs_2013}. Even more remarkably, long range order can be not only destroyed but also created by exciting samples with light. For instance, the dynamic emergence of transient spin-density wave~\cite{kim_ultrafast_2012}, charge-density wave~\cite{singer_enhancement_2015}, as well as superconducting order~\cite{fausti_light-induced_2011, mankowsky_nonlinear_2014, hu_optically_2014, kaiser_optically_2014, mitrano_possible_2016} has been demonstrated. These observations lead to the very fundamental theoretical questions of the origin of the light induced order, including: What is the mechanism for the emergence of transient collective behavior? What determines the lifetime of transient ordered states? How robust and universal are the observed phenomena? Answers to these questions may hold the key to a novel route for achieving ordered many-body states by periodic driving as opposed to cooling; a subject that has attracted considerable theoretical attention recently~\cite{subedi_theory_2014, kemper_amplitude_2014, goldstein_photoinduced_2015, raines_enhancement_2015, sentef_theory_2016, orenstein_terahertz_2015, murakami_multiple_2015}.

\begin{figure*}
  \centering 
  \includegraphics[width=.98\textwidth]{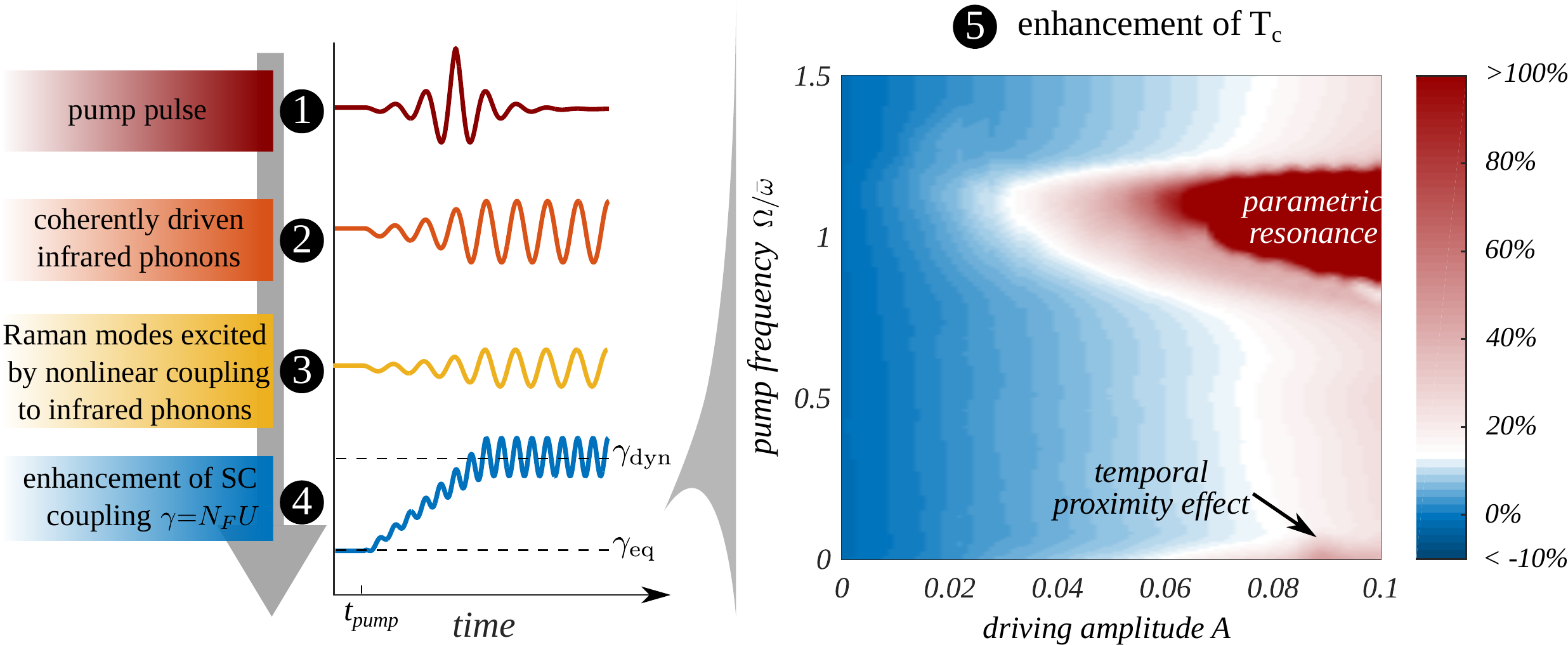}
  \caption{ \textbf{Dynamical enhancement of the superconducting transition temperature.}   A schematic representation of the physical processes leading to light-induced superconductivity: (1) The pump pulse couples coherently to (2) an infrared-active phonon mode which in turn (3) via nonlinear interactions drives Raman phonons that are responsible for superconducting pairing. The non-equilibrium occupation of the Raman phonons (4) universally enhances the superconducting coupling strength $\gamma$, which is a product of the density of states at the Fermi level $N_F$ and the induced attractive interaction between the electrons $U$, and hence (5) increases the transition temperature $T_c$ of the superconducting state. We calculate the relative enhancement of $T_c$ compared to equilibrium $(T_c-T_{c,\text{eq}})/T_{c,\text{eq}}$ by taking into account the competition between dynamical Cooper pair-formation and Cooper pair-breaking processes, as a function of the pump frequency $\Omega/\bar \omega$ and the driving amplitude $A$. The data is evaluated for linearly dispersing phonons with mean frequency $\bar \omega$, relative spread $\Delta \omega/\bar \omega =0.2$, and negative quartic couplings of type II between the Raman and infrared-active modes; \tab{tab:phonon}. Moreover, the electron-phonon interaction strength is chosen to give an equilibrium effective attractive interaction $U/W = 1/8$ that is weak compared to the bare electronic bandwidth $W$. The static renormalization of Raman modes leads to the uniform increase of $T_c$ with increasing driving amplitude $A$, the squeezed phonon state manifests in the strong enhancement near parametric resonance $\Omega \sim \bar \omega$, and the temporal proximity effect dominates near $\Omega/\bar \omega \sim 0$. 
  } 
  \label{fig:enh}
\end{figure*}
\begin{table}[b]
\centering
\caption{\textbf{Types of the phonon nonlinearities and their static and non-equilibrium effects on the system.} The renormalization of the parameters leads to an effectively static Hamiltonian, while phonon squeezing and the periodic Floquet modulation of system parameters are purely dynamical.}
\label{tab:phonon}
\begin{tabular}{llccc}
\hline
  & \multicolumn{1}{c}{}  & \begin{tabular}[c]{@{}c@{}}\;\;static renorm.\;\;\\ of parameters\end{tabular} & \begin{tabular}[c]{@{}c@{}}\;\;dynamical\;\;\\ squeezing\end{tabular} &  \begin{tabular}[c]{@{}c@{}}\;\;periodic \\ Floquet\end{tabular} \\\hline
\multicolumn{1}{l}{I:}   & $(Q^\text{IR}_0)^2 Q_0^\text{R}$  & \checkmark & $\times$  & \checkmark \\ 
\multicolumn{1}{l}{II:}  & $(Q^\text{IR}_0)^2 Q_k^\text{R} Q_{-k}^\text{R}$  & \checkmark & \checkmark & \checkmark \\ 
\multicolumn{1}{l}{III:} & $Q^\text{IR}_0 Q_k^\text{R} Q_{-k}^\text{R}$ & $\times$ & \checkmark & \checkmark \\ \hline
\end{tabular}
\end{table}

We propose a general mechanism for making a normal conducting metal unstable toward Cooper-pair formation by irradiation with light. The key ingredient is the nonlinear coupling between optically active infrared phonons and the Raman phonons that mediate electron-electron attraction, responsible for superconductivity (\fig{fig:enh}). Depending on the form of the nonlinearity (\tab{tab:phonon}), several effects can arise. These include the parameter renormalization of the \emph{time-averaged} Hamiltonian, the \emph{dynamic} excitation of the Raman phonons into {\em squeezed quantum states}, which can have significantly enhanced coupling to electrons, and, finally, the \emph{periodic} modulation of system parameters leading to Floquet states. Of these effects, we find that phonon squeezing universally leads to an enhancement of $T_c$, by potentially a large factor~\footnote{Squeezed states of phonons have also been studied in the context of optomechanical systems as a mechanism to enhance nonlinear couplings~\cite{lemonde_enhanced_2016}}. The periodic modulation of the system parameters (Floquet) can also enhance Cooper pairing, via a superconducting proximity effect in time rather than space. 
By contrast, the static renormalization of the system parameters can, depending on material-specific details, either enhance or suppress $T_c$.

Driving phonons into a highly excited state, unfortunately, also leads to an increased electron-phonon scattering rate, which {\em weakens} Cooper pairing. We analyze the competition between the enhanced Cooper pair formation and Cooper pair breaking and show that the enhancement of pairing can dominate in a broad parameter range resulting in signatures of superconductivity that appear at higher temperatures compared to equilibrium. We note that the predicted enhancement of $T_c$ should be understood as a transient phenomenon, since inelastic scattering of electrons with excited phonons will eventually heat the system and destroy the superconducting order. Even though our study is motivated by specific experiments, the minimal models that we introduce and study are intended to elucidate the qualitative origins of these effects, rather than to provide detailed material-specific predictions.

Furthermore, the proposed mechanism can be readily applied to many other types of long-range order such as charge-density and spin-density waves, by performing the instability analysis in the appropriate channels.

\section{Role of the phonon nonlinearity}

We now describe the consequences of different types  of optically-accessible phonon nonlinearities.
The  pump pulse directly couples to infrared-active phonon modes, which have a finite dipole moment. Since the photon momentum is negligible compared to the reciprocal lattice vector, the drive  creates a coherent phonon state at zero momentum, $Q^\text{IR}_{q=0}(t)=\mathcal{E} \cos \Omega t$, where $\Omega$ is the drive frequency and $\mathcal{E}$ is proportional to the drive amplitude. In the presence of phonon nonlinearities, the driven infrared-active phonon mode couples to Raman modes $Q^\text{R}_q$ of the crystal~\cite{forst_nonlinear_2011, mankowsky_nonlinear_2014, subedi_theory_2014, forst_mode-selective_2015}, which in turn can couple to the conduction electrons. 

There are three leading types of phonon nonlinearities which can have static and dynamic effects (\tab{tab:phonon}). A static renormalization of the Hamiltonian parameters arises from phonon nonlinearities that involve even powers of $Q_0^\text{IR}$, since these terms have finite time averages. For phonon nonlinearities of type I (\tab{tab:phonon}), this  leads to a static {\em displacement} of the zero-momentum Raman phonon while for nonlinearities of type II the {\em frequency} of the Raman phonons is modified, \textit{cf.} App.~\ref{sec:zmp}. Both affect the superconducting instability, causing either enhancement or suppression; the analysis can be done with a standard equilibrium Migdal-Eliashberg formalism~\cite{mahan_many_2000}. Here we will focus, however, primarily on the {\em dynamical} effects. There are two types of dynamical effects that can be distinguished:  First, a simple modulation of system parameters, which makes system {\em instantaneously} more or less superconducting, and, second, dynamical squeezing of phonons, which is an explicitly quantum effect. Notably, both dynamical effects lead to an increased superconducting instability temperature.

\section{A minimal electron-phonon model}

We illustrate the aforementioned effects by using a model with Fr\"ohlich-type electron-phonon interactions that couple with strength $g_k$ to the displacement of the Raman phonons $Q_k^\text{R}$ ($P_k^\text{R}$ is the conjugate momentum) to the local electron density $n_{i\sigma}$: 
\begin{align}
\hat H_\text{el-ph} &=  -J_0\sum_{\langle ij\rangle,\sigma} c_{i\sigma}^\dag c_{j\sigma} + \sum_k ( P_k^\text{R} P_{-k}^\text{R} + \omega_k^2 Q_k^\text{R} Q_{-k}^\text{R}) \nonumber\\&+ \sum_{ik\sigma} \sqrt{\frac{2\omega_k}{V}} g_k e^{ikr_i} \, Q_k^\text{R} \, n_{i\sigma}^\nag. 
\label{eq:hep}
\end{align}
We consider dispersive optical phonons $\omega_k$ with mean frequency $\bar \omega$ and spread $\Delta \omega$. Here, $J_0$ is the bare electron hopping matrix element and $V$ the volume of the system. Further, we assume a quartic nonlinearitiy of type II. The phonon drive term, introduced for $t>0$, reads
\begin{equation}
 \hat H_\text{drv}(t)=-\sum_k \omega_k^2 A_k (1+ \cos 2\Omega t) Q_k^\text{R} Q_{-k}^\text{R},
 \label{eq:phdrv}
\end{equation}
where $A_k=-{\Lambda_k \mathcal{E}^2}/{2 \omega_k^2}$ and $\mathcal{E}$ is the amplitude of the driven infrared-active phonon.

The effect of the drive is twofold. First, there is a static contribution, which renormalizes Raman phonon frequencies. For negative nonlinearities, $\Lambda_k<0$, phonons are softened by $\omega_k^2(1-A_k)$. The mode softening suppresses the electron tunneling, increases the density of states and thus the overall pairing strength. Second, the time dependent part of the drive \eqw{eq:phdrv} realizes a parametrically driven oscillator which dynamically generates phonon squeezing correlations and can strongly suppress the electron tunneling matrix element~\footnote{Even without drive, i.e. in equilibrium, the electron tunneling is suppressed by polaronic dressing. While hopping on the lattice, the electron distorts it and slows down. The hopping matrix element is suppressed by the Franck-Condon overlap: $J_\text{eq}=J_0 \exp[{-{g^2}/{\omega^2}}]$. Out of equilibrium, the polaronic effect is enhanced due to dressing the by {squeezed} states of phonons.}. While the mode softening effect on $T_c$ depends on the sign of the nonlinearity, the dynamical generation of squeezing correlations \emph{universally} leads to enhancment of $T_c$.

\textit{Effective Hamiltonian.---}We will now derive an effective electronic Hamiltonian by integrating out phonons. This is analogous to the standard derivation of the Bardeen-Cooper-Schrieffer (BCS) Hamiltonian starting from the Fr\"ohlich model, however, with the phonons being driven strongly out of equilibrium. As a consequence, the Hamiltonian will be explicitly time-dependent. In addition, it will contain residual electron-phonon scattering that contribute to the electron scattering rate.

We first perform the Lang-Firsov transformation (App.~\ref{sec:fmp}), which eliminates the electron-phonon interaction term in \eq{eq:hep} but introduces an effective electron-electron interaction and dresses the electron tunneling with phonons. This dressing, which depends on the phonon squeezing, suppresses the electron tunneling and modulates it in time.
We (i) take into account the softening of the Raman modes by a static renormalization of the phonon coordinates and (ii) treat the dynamic excitation of finite-momentum phonons by transforming the system into a rotating frame. Since we are considering optical phonons, we assume their thermal occupation to be negligible before the drive was switched on. Tracing out the  phonons, leads to the dressed electron tunneling matrix element $\hat H_\text{kin} \to J(t) \sum_{ij\sigma}  c_{i\sigma}^\dag c_{j\sigma}^\nag$. (iii) We compute the rate of non-equilibrium Cooper pair breaking processes resulting from dynamical excitations of the Fermi sea induced by the drive and take into account the competition between the enhanced pair formation and pair breaking. This treatment is justified by showing \textit{a posteriori} that pair formation dominates pair breaking.

Solving the full problem numerically we find that $J (t)$ oscillates with $2 \Omega$ around its mean value 
\begin{equation}                                                                                                                                                                                                                                                                                                                                                                                                                                                                                                                                                                                                                                                                                                                                                                                                                                                                                                                                                                                                                                                                                                                                                                                                                                                                                                                                                                                 
\langle J(t) \rangle = J_\text{eq} e^{-\zeta},
 \label{eq:Jtt}
\end{equation}
which is suppressed by mode softening and phonon squeezing, as parametrized by $\zeta$. Near parametric resonance the squeezing correlations increase in time leading to a decrease in $J(t)$ and an increase in the amplitude of its oscillations. Since we are studying transient phenomena, we compute the average over the first ten driving cycles to extract the effective electron tunneling. We take the electron-phonon coupling $g_k$ such that $g_k^2/\omega_k = \text{const.}$, which leads to a local Hubbard-type electron-electron interaction $U$. This assumption is reasonable for phonons with wavevectors above the Thomas-Fermi screening length~\cite{mahan_many_2000}. We emphasize however that this assumption is not crucial for our analysis. For technical convenience, we transform the oscillatory part from the kinetic to the interaction term by rescaling time, yielding the effective time-dependent Hamiltonian
\begin{align}
 \tilde H(t) &= J_\text{eq} e^{-\zeta} \sum_{ij} c_{i\sigma}^\dag c_{j\sigma}^\nag -U (1+\mathcal{A}\cos 2 \Omega t) \sum_i n_{i\uparrow} n_{i\downarrow} \nonumber\\&+ \hat H_\text{el-ph scatt.} ,
 \label{eq:H}
\end{align}
where $\hat H_\text{el-ph scatt.}$ represent electron-phonon scattering terms that vanish upon tracing out the phonons in the rotating frame. The drive, \eq{eq:phdrv}, has thus several effects: (i) a suppression of the electron tunneling by a factor $\exp[-\zeta]$, (ii) a dynamic Floquet contribution from modulating the interaction energy by $\mathcal{A} \cos 2 \Omega t$, and (iii) an enhancement of the electron scattering due to the non-equilibrium phonon occupation. 

\textit{Dynamical Cooper instability.---}We study the dynamical Cooper instability toward pair formation in Hamiltonian \eqw{eq:H} by combining the BCS approach~\cite{bardeen_theory_1957} with a Floquet analysis~\cite{bukov_universal_2015}. We also take into account the finite electron lifetime $\tau$ due to the non-equilibrium phonon occupation, by introducing an imaginary self-energy correction $i/\tau$ [calculated in (App.~\ref{sec:fmp}) using Floquet Fermi's Golden rule]. In contrast to elastic scattering on a time-reversal symmetry preserving potential, where Anderson's theorem~\cite{anderson_theory_1959,abrikosov_theory_1959} shows that the thermodynamics of a superconductor remains unchanged, the inelastic scattering processes arising here indeed alter the superconducting properties. 
\begin{figure}[t!] 
  \centering 
  \includegraphics[width=.46\textwidth]{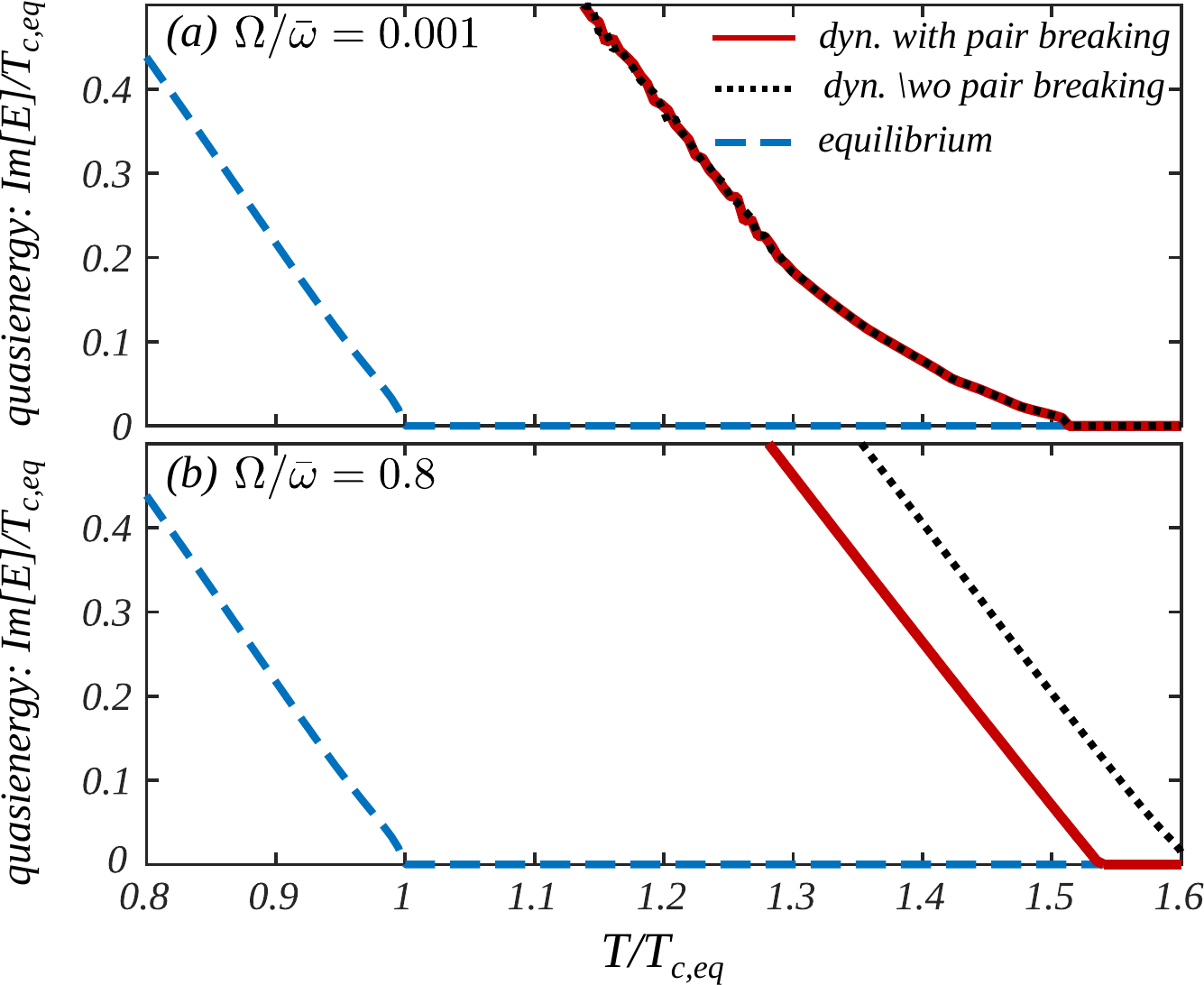}
  \caption{ \textbf{Dynamical Cooper instability.} The dynamical Cooper instability evaluated within the Floquet BCS theory \emph{including} non-equilibrium pair-breaking processes, red solid line, is compared to the one without pair breaking, black dotted line, and the BCS solution for the equilibrium problem, blue dashed line.
  Data is evaluated for mean phonon frequency $\bar \omega/W =1/8$ and effective attractive interactions $U/W=1/8$ that are weak compared to the bare electronic bandwidth $W$, a phonon frequency spread of $\Delta \omega=0.2\bar \omega$, driving strength $A=0.1$, and driving frequency \fc{a} $\Omega/\bar \omega=0.001$ and \fc{a} $\Omega/\bar \omega=0.8$. The non-equilibrium Cooper pair-formation rate dominates over the pair breaking, hence, leading to an enhanced transition temperature. }
  \label{fig:gap}
\end{figure}

To carry out the BCS Floquet analysis of the pairing instability, we use the equations of motion technique~\cite{anderson_random-phase_1958}. We introduce infinitesimal pairing amplitudes $a_k = \langle c_{k\uparrow} c_{-k\downarrow} \rangle$ and determine whether the system is stable or unstable to the growth of $a_k$~\cite{abrikosov_methods_1975, pekker_competition_2011} by finding the eigenfequencies of the corresponding collective mode. We decompose $a_k(t)$ into Floquet modes
 $a_k(t) \sim e^{-i E t} \sum_{n=-\infty}^{\infty}  a_{kn} e^{2 in\Omega t}$,
where $E$ is the energy which has to be determined self-consistently from the Floquet BCS gap equation (App.~\ref{app:bcs}) 
\begin{subequations}
\begin{align}
0&=(U^{-1}+F_n )\Delta_n + \frac{\mathcal{A}}{2}F_n(\Delta_{n-1}+\Delta_{n+1}) \\
F_n&=\frac{1}{V}\sum_k \frac{1-2n_k}{E+2 n\Omega - 2 (\epsilon_k + i/\tau -\mu - U \rho) } .%
\end{align}%
\label{eq:gap}%
\end{subequations}
The instability of the system manifests in the appearance of an eigenmode with negative imaginary part of $E$ which we obtain by searching for the zeros of the determinant of \eqw{eq:gap} in the complex plane of $E$~\cite{abrikosov_methods_1975, pekker_competition_2011}. Here, $\Delta_n=\frac{U}{V}\sum_k a_{kn}^*$ are the Floquet harmonics of the gap, $\rho$ is the electron density of a single spin-component, $\mu$ the chemical potential, $n_k$ the Fermi-Dirac distribution of the electrons determined by the temperature of the undriven system, and $\epsilon_k$ the electron dispersion.  The assumed thermal state of the electrons is justified in the weak-coupling regime and since we are only interested in transient effects that follow the drive~\cite{bauer_dynamical_2015}. We determine the critical $T_c$ by locating the highest temperature at which we find an unstable solution of the Floquet BCS gap equations \eqw{eq:gap}. Equations with similar structure are obtained for spatially inhomogeneous superconductors~\cite{martin_enhancement_2005}.

The high and low frequency limits of the Floquet BCS gap equations \eqw{eq:gap} can be understood from perturbative arguments. In the high frequency limit, $\Omega \to \infty$, we use a Magnus expansion to derive the stroboscopic Floquet Hamiltonian~\cite{bukov_universal_2015}. To zeroth order in $1/\Omega$, the Floquet Hamiltonian is given by the time averaged Hamiltonian. Thus for the harmonic drive in \eq{eq:H}, the contribution $\sim \mathcal{A}$ drops out and we obtain an equilibrium BCS gap equation with interaction $U$ and the  reduction of the electron tunneling $J_\text{eq} \exp[-\zeta]$ due to the squeezed phonon state. Enhancement of $T_c$ due to the suppression of the effective electron bandwidth has also been suggested in a model without phonons~\cite{raines_enhancement_2015}.

In the low frequency limit $\Omega \to 0$, the core of the gap equations \eqw{eq:gap} barely depends on the Floquet index $n$ leading to
 $[U^{-1}+F (1+\mathcal{A})]\Delta = 0$.
The maximally enhanced transition temperature $T_c$ is thus determined by an equilibrium BCS gap equation with $U (1+\mathcal{A})$ which is the largest instantaneous attractive interaction. Hence, in the slow drive limit, the pairing induced by the strongest instantaneous interaction dominates the Cooper pair formation which can be interpreted as a superconducting proximity effect in time rather than in space. 

We verify these perturbative predictions by numerically solving the Floquet BCS gap equations \eqw{eq:gap} on a two-dimensional square lattice away from half-filling.  The convergence of the results with system size and number of Floquet bands is checked. We choose linearly dispersive phonons with mean frequency $\bar \omega/W=1/8$, interaction energy $U/W=1/8$ weak compared to the bare electron bandwidth $W$, and set the width of the phonon branch to $\Delta \omega = 0.2 \bar \omega$; a typical value for optical phonons. We first solve the driven phonon problem from which we determine the average suppression of the electron tunneling $e^{-\zeta}$ and the effective amplitude $\mathcal{A}$ of the oscillations in the interaction. Then we solve the Floquet BCS problem. In \fig{fig:gap} we show the dynamical Cooper instability including pair-breaking processes for driving $\Omega/\bar\omega=\{0.001,\ 0.8\}$ and $A_k=A=0.1$ constant for all $k$, red solid line, and compare them to the instability without pair breaking, black dotted line, and the equilibrium solution, blue dashed line. 

The reduction of the superconducting transition temperature due to the finite electron lifetime is of the Abrikosov-Gorkov form $T_c=T_{c,0}-1/\tau$, where $T_c$ is the dynamical transition temperature including pair-breaking and $T_{c,0}$ the one without pair-breaking. We find an enhancement of the dynamical $T_c$ compared to equilibrium, \fig{fig:gap}, and hence that the pair-formation rate dominates the pair-breaking rate. In the high-frequency limit $\Omega/\bar \omega = 0.8$, \fc{b}, the enhancement results mainly from the efficient suppression of the electron tunneling near parametric resonance $\Omega \sim \bar \omega$ and in the slow drive limit $\Omega \to 0$, \fc{a}, from a combination of mode softening and the temporal proximity effect. The relative enhancement of $T_c$, taking into account pair-breaking processes, is shown in \fig{fig:enh} for an extended range of drive amplitudes $A$ and frequencies $\Omega/\bar \omega$. Softening of the Raman modes leads to the uniform increase of $T_c$ with driving amplitude for all values of the pump frequency, while the effect of phonon squeezing is most prominant near parametric resonance $\Omega \sim \bar \omega$ and can dominate mode softening by about an order of magnitude. As a result, the enhancement of $T_c$ displays an intricate non-monotonic dependence on the driving frequency.

\section{Experimental implications}

Even though we are not studying a specific material from first principles, let us estimate the typical order of magnitude of the discussed effects for a recent experiment in which transient non-equilibrium superconductivity has been explored in \KC fullerides~\cite{mitrano_possible_2016}. Typical energy scales in \KC are the following~\cite{gunnarsson_superconductivity_1997}: bare bandwidth $W \sim 0.6\text{eV}$, intramolecular phonon frequencies $\omega \sim 0.03 - 0.2 \text{eV}$, and interaction $U=g^2/\omega \sim 0.1\text{eV}$. Thus, the parameters we chose are representative for \KC. 
In the experiment~\cite{mitrano_possible_2016} the driving frequency $\Omega$ ranges from a tenth to a third of the bandwidth. Translating to our scenario, the main enhancement of $T_c$ in this experiment should result from a combination of the \emph{static} mode softening and the \emph{dynamic} phonon squeezing.

Our calculations suggest that the increase of $T_c$ is related to a dynamical enhancement of the effective electron mass, \eq{eq:Jtt}. Such a dynamical renormalization of the electronic dispersion can, for instance, be determined experimentally by time-resolved ARPES measurements, see e.g. \cite{schmitt_transient_2008, cortes_momentum-resolved_2011, graf_nodal_2011, smallwood_tracking_2012}. Contributions from mode softening and phonon squeezing can, in principle, be experimentally distinguished by the driving-strength dependence of the mass renormalization, that we predict to be linear in the former and quadratic in the latter case.

\section{Summary and outlook}
We studied the competition between the increased Cooper pair formation and Cooper pair breaking in strongly-driven electron-phonon systems and found that the latter need not to inhibit the dynamic enhancement of $T_c$. Even though we focused on superconducting instabilities, the proposed mechanism for achieving a larger coupling constant is generic and can be directly applied to other forms of long-range order such as spin or charge density waves.

The present analysis addresses the transient dynamics of the system. At long times, the inelastic scattering of electrons from excited phonons will increase their temperature and will tend to destroy superconducting order, as seen in the experiments. At intermediate times, heating of the electrons can give rise to nonequilibrium distribution functions which potentially enhances superconducting coherence as experimentally seen by microwave irradiation of superconducting samples~\cite{wyatt_microwave-enhanced_1966, dayem_behavior_1967, eliashberg_1970, ivlev_nonequilibrium_1973, klapwijk_radiation-stimulated_1977, schmid_stability_1977}. 
The analysis of complete thermalization due to feedback effects in a fully self-consistent and conserving calculation is an exciting future direction.

\begin{acknowledgments}
We thank E. Berg, A. Cavalleri, I. Cirac, U. Eckern, D. Fausti, A. Georges, S. Gopalakrishnan, B.I. Halperin, A. Imamoglu, S. Kaiser, C. Kollath, M. Norman, T. Shi for useful discussions. We acknowledge support from the Technical University of Munich - Institute for Advanced Study, funded by the German Excellence Initiative and the European Union FP7 under grant agreement 291763, DFG grant No. KN 1254/1-1, Harvard-MIT CUA, NSF grant No. DMR-1308435 and DMR-1410435, AFOSR Quantum Simulation MURI, the ARO-MURI on Atomtronics, Humboldt foundation, Dr.~Max R\"ossler, the Walter Haefner Foundation, the ETH Foundation, the Simons foundation, as well as the Institute for Quantum Information and Matter, an NSF Physics Frontiers Center with support of the Gordon and Betty Moore Foundation. This work was supported by the U.S. Department of Energy, Office of Science, Materials Sciences and Engineering Division.
\end{acknowledgments}

\appendix

\begin{onecolumngrid}


\section{Type I nonlinearity: Uniform lattice displacement \label{sec:zmp}}

A nonlinear phonon coupling of form I, $\Lambda (Q^\text{IR}_0)^2 Q^\text{R}_{0}$ (\textit{cf.} Tab. 1 in the main text), acts as a classical force on $Q^\text{R}_0$ proportional to $(Q^\text{IR}_0)^2$~\cite{forst_nonlinear_2011, mankowsky_nonlinear_2014, subedi_theory_2014, forst_mode-selective_2015}. The drive $(Q^\text{IR}_0)^2 = \mathcal{E}^2 \cos^2 \Omega t = \frac{\mathcal{E}^2}{2}(1+\cos 2 \Omega t)$ can be separated in a static and an oscillatory contribution. We take into account a simple electron-phonon model in which the phonons modulate electron hopping processes $-(J_0+g\, Q^\text{R}_0) \sum_{\langle ij \rangle,\sigma} \, c_{i\sigma}^\dag c_{j\sigma}^\nag$ ($g$ is the electron-phonon coupling and $c_{i\sigma}^\dag$ the electron creation operator) as for example in a Su-Schrieffer-Heeger (SSH) model~\cite{heeger_solitons_1988}
\begin{equation}
  H_\text{SSH} = -(J_0+g\, Q^\text{R}_0) \sum_{\langle ij \rangle,\sigma} \, c_{i\sigma}^\dag c_{j\sigma}^\nag+\frac12 \sum_k (P^\text{R}_k P^\text{R}_{-k} + \omega_k^2 Q^\text{R}_k Q^\text{R}_{-k}) + \Lambda (Q_0^\text{IR})^2 Q^\text{R}_0.
\end{equation}
The infrared phonon is coherently driven $Q_0^\text{IR} = \mathcal{E} \cos \Omega t$. Neglecting the feedback of the electrons on the phonons, we find for the phonon equation of motion
\begin{equation}
 \ddot Q_0^\text{R}+\omega_0^2 Q_0^\text{R} = \frac{\Lambda \mathcal{E}^2}{2}(1+\cos 2\Omega t).
\end{equation}
This equation can be solved analytically with $Q_0^\text{R}(t)=\tilde Q + \delta Q \cos 2\Omega t$, where $\tilde Q = \frac{\Lambda \mathcal{E}^2}{2\omega_0^2}$ and $\delta Q = \frac{\Lambda \mathcal{E}^2}{2(\omega_0^2-4\Omega^2)}$. The phonon displacement thus oscillates with twice the pump frequency $2\Omega$ around a mean value $\tilde Q$. Plugging this into the SSH Hamiltonian, we find
\begin{align}
  H_\text{SSH} &=  -[J_0 +  g(\tilde Q + \delta Q \cos 2\Omega t)] \sum_{\langle ij \rangle,\sigma} c_{i\sigma}^\dag c_{j\sigma}+\frac12 \sum_{k\neq0} (P^\text{R}_k P^\text{R}_{-k} + \omega_k^2 Q^\text{R}_k Q^\text{R}_{-k}).
\end{align}
The finite displacement of the lattice along the coordinates of the Raman mode renormalizes the electron tunneling by a term $\propto g \Lambda \mathcal{E}^2$. The effective electronic bandwidth is reduced when $g\Lambda < 0$; a condition that is material specific. Reducing the bandwidth results in a higher density of states and hence an enhanced superconducting transition temperature $T_c$. For $g\Lambda>0$ contrary is the case. These general considerations may account for the physical origin of the enhancement of the transition temperature which has been seen in \textit{ab initio} calculations for specific materials~\cite{mankowsky_nonlinear_2014}. On top of the enhancement due to the time-averaged displacement of the Raman phonon mode, the strong oscillations at $2\Omega$ can give rise to an additional dynamic enhancement of $T_c$ by the temporal proximity effect as we discuss in this work.

\section{Type II and III nonlinearities: Finite-momentum phonon excitations \label{sec:fmp}}

An alternative form of the nonlinear phonon interaction contains pairs of Raman phonons at finite but opposite momenta. Here, we analyze both the quartic type II and the cubic type III nonlinearities (\textit{cf.} Tab. 1 in the main text for their classification).

\textit{Type II: Quartic nonlinearities.---}The lowest order nonlinearity in centro-symmetric crystals containing pairs of Raman phonons is of the quartic form  $\Lambda_k (Q^\text{IR}_{0})^2 Q^\text{R}_{k} Q^\text{R}_{-k}$ (type II). In this case, the coherently driven infrared phonon excites Raman phonons in pairs and with opposite finite momenta, leading to quantum correlations between $k$ and $-k$ modes. These states are referred to as \emph{squeezed states} in quantum optics~\cite{walls_quantum_2008}. The driven phonon Hamiltonian has the form  
\begin{align}
\hat H_\text{ph}=\sum_k ( P_k^\text{R} P_{-k}^\text{R} + \omega_k^2 Q_k^\text{R} Q_{-k}^\text{R}) -\sum_k \omega_k^2 A_k(1+\cos 2 \Omega t) Q_k^\text{R} Q_{-k}^\text{R} = \hat H_{\text{ph},0} + \hat H_\text{drv.},
\end{align} 
with $A_k=-\Lambda_k \mathcal{E}^2/2\omega_k^2$, which describes a parametric oscillator with resonance condition $\Omega = \omega_k \sqrt{1-A_k}$.

The drive can be again separated into a static and an oscillatory contribution with the implications of the static displacement being crucially dependent on the sign of the nonlinearity $\Lambda_k$. For $\Lambda_k<0$ the frequencies of the Raman modes $\omega_k$ are softened and hence the effective electron-electron interaction, which is typically of the form $g^2_k/\omega_k$, is increased, while for $\Lambda_k>0$ the opposite is the case. By contrast, we find that irrespective of the sign of the couplings, the oscillatory term squeezes the Raman modes which results in a polaronic suppression of the electron tunneling matrix element compared to equilibrium and hence in an enhanced density of states, \figc{fig:EnhDiag_univ}{a}. Such an enhancement in the density of states in turn increase the superconducting coupling constant, enabling dynamic Cooper pair formation at higher temperatures where equilibrium Cooper pairing would be impossible.

\textit{Type III: Cubic nonlinearities.---}In non-centrosymmetric crystals, the lowest order phonon nonlinearity is of the cubic form III, $Q^\text{IR}_{0} Q^\text{R}_{k} Q^\text{R}_{-k}$, as the zero-momentum infrared active phonon mode can linearly couple to the displacement of the Raman modes. Coupling to pairs of Raman modes at opposite momentum gives rise to phonon squeezing that leads to an enhanced $T_c$. Hence, phonon squeezing is largely insensitive to the microscopic details of the phonon nonlinearities and can occur in a variety of nonlinear couplings. The effective phonon Hamiltonian arising from such a nonlinearity is
\begin{align}
\hat H_\text{ph}=\sum_k P_k^\text{R} P_{-k}^\text{R} + \omega_k^2(1+A_k\cos \Omega t) Q_k^\text{R} Q_{-k}^\text{R} ,
\end{align} 
with $A_k=\Lambda_k \mathcal{E}$, which also describes a parametric oscillator but with resonance condition $\Omega=2\omega_k$. In the case of cubic interactions, direct mode softening cannot occur because the the time average of $Q_0^\text{IR}$ vanishes. Nonetheless, the dynamic generation of squeezing near the parametric resonance is universally present also for cubic nonlinearities and qualitatively similar enhancement diagrams can be also obtained in that case, \textit{cf.}~\figc{fig:EnhDiag_univ}{b}.\\

\begin{figure*}
        \centering
        \includegraphics[width=0.98\textwidth]{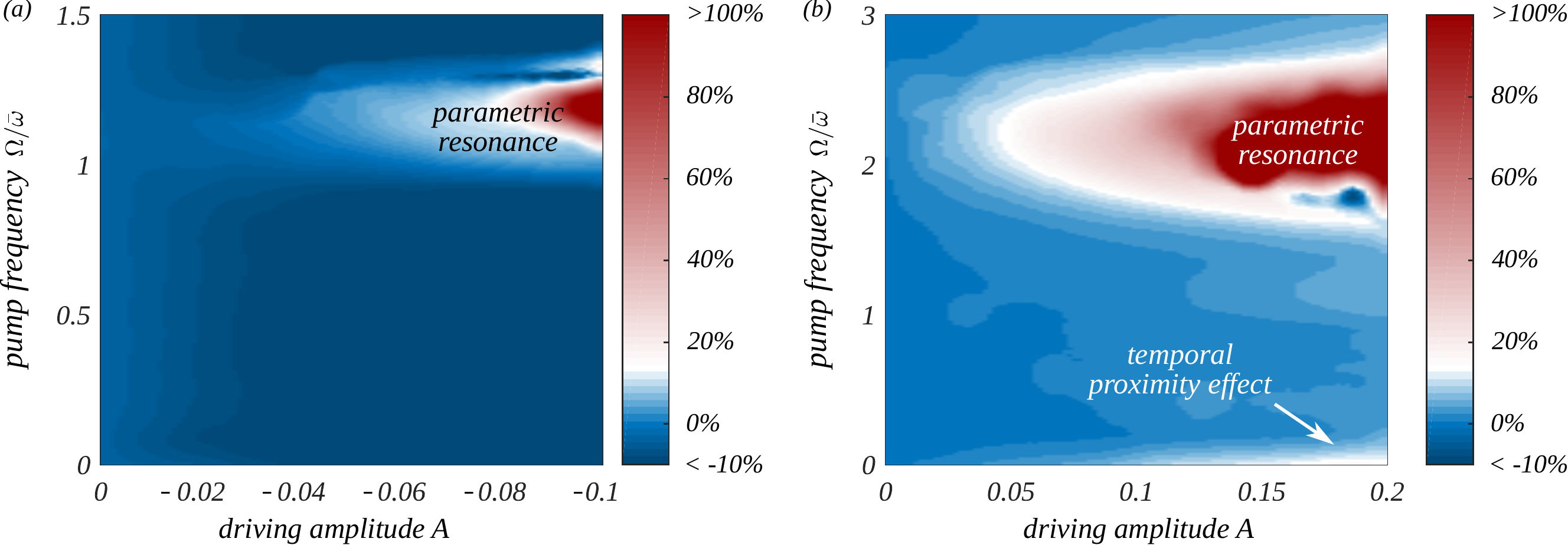}
        \caption{\textbf{Phonon squeezing as a universal mechanism for the enhancement of $T_c$}. The relative change of the superconducting transition temperature is shown for \fc{a} quartic type II phonon nonlinearities with positive $\Lambda_k>0$ and \fc{b} cubic type III phonon nonlinearities, taking into account the competition between Cooper pair formation and pair breaking. \fc{a} Quartic phonon nonlinearities that couple to finite momentum Raman modes have two distinguished effects on the bandwidth renormalization: (i) the static displacement of the Raman modes and (ii) phonon squeezing. For positive nonlinearities, the static displacement hardens the frequency of the Raman phonons. This leads to the uniform suppression of $T_c$ with increasing magnitude of the driving amplitude $A$. However, near parametric resonance $\Omega \sim \bar \omega$, the phonon squeezing dominates over the static displacement, leading to the universal enhancement of $T_c$. \fc{b} Cubic nonlinearities exhibit only the squeezing mechanism, and hence irrespectively of the sign of the coupling lead to an enhanced transition temperature. Furthermore, the shape of the enhancement diagram is similar to the one for quartic interaction with the main difference that in the case of quartic nonlinearities the resonance condition is $\Omega \sim \bar \omega$ while in the cubic case it is $\Omega \sim 2\bar \omega$. The data is evaluated for $\bar \omega/W=1/8$, $U/W=1/8$, $\Delta \omega = 0.2 \bar \omega$. 
        }
        \label{fig:EnhDiag_univ} 
\end{figure*}

For studying phonon nonlinearities which generate squeezed states, we consider the Holstein model with Fr\"ohlich electron-phonon interactions where the phonon displacement couples to the charge density
\begin{equation}
H_\text{el-ph} = -J_0 \sum_{ij\sigma} c_{i\sigma}^\dag c_{j\sigma}^\nag + \sum_k \omega_k b_k^\dag b_k + \frac{1}{\sqrt{V}}\sum_{ik} g_k  (b_k^\nag + b_{-k}^\dag)e^{ikr_i}n_{i\sigma}.
\label{eq:Hol}
\end{equation}
We expect our results to be insensitive to the fine details of the model and expect for example similar results for the SSH model.

We pursue the following strategy:
\begin{enumerate}
 \item Using a Lang-Firsov transformation we remove the electron-phonon interaction.
 \item We determine the unitary transformation $W$ that transforms the system into a rotating frame in which the phonon driving is absent. This leads to a driven Floquet BCS problem.
 \item We estimate the dynamic Cooper pair breaking rate due to the non-equilibrium scattering of phonons and due to oscillations in the microscopic interaction parameter using Floquet Fermi's Golden rule. The Cooper pair-breaking rate is self-consistently taken into account by adding an imaginary self-energy correction to the Floquet BCS equations.
 \item Solving the driven Floquet BCS problem, we determine the enhancement of $T_c$ relatively to the undriven one.
\end{enumerate}

\subsection{Lang-Firsov transformation}

Considering the Holstein model~\eqw{eq:Hol}, we introduce operators $\tilde b_k$ and $\tilde b_{-k}^\dag$ which diagonalize the static phonon Hamiltonian $\sum_k  P_k^\text{R} P_{-k}^\text{R} + \omega_k^2(1-A_k) Q_k^\text{R} Q_{-k}^\text{R}$. These operators are related to the $b_k^\nag$, $b_{-k}^\dag$ operators which diagonalize $H_{\text{ph},0}$ e.g. by $(b_{-k}^\dag-b_k^\nag)=(1-A_q)^\frac{1}{4} (\tilde b_{-k}^\dag-\tilde b_k^\nag)$. The electron-phonon interaction in \eq{eq:Hol} can be removed by a Lang-Firsov transformation
\begin{equation}
 H_\text{el-ph} \to e^{S} H_\text{el-ph} e^{-S} \qquad \text{with} \qquad S= - \frac{1}{\sqrt{V}}\sum_{qj\sigma} \frac{g_q}{\omega_q \sqrt{1-A_q}} e^{iqr_j} (\tilde b_q^\nag-\tilde b_{-q}^\dag)n_{j\sigma}.
\end{equation}
Applying it to the Hamiltonian and switching on the drive we obtain
\begin{equation}
 H  = -\sum_{ij\sigma} J_{ij}c_{i\sigma}^\dag c_{j\sigma}^\nag+\sum_{ij \sigma \sigma'} U_{ij} n_{i\sigma} n_{j\sigma'} + \sum_k  P_k^\text{R} P_{-k}^\text{R} + \omega_k^2(1-A_k-A_k\cos 2 \Omega t) Q_k^\text{R} Q_{-k}^\text{R}. 
 \label{eq:driveneph}
\end{equation}
with dressed tunneling matrix element $J_{ij}=J_0 e^{- \frac{1}{\sqrt{V}} \sum_k \frac{g_k}{\omega_k(1-A_k)^{{3}/{4}}} (e^{ikr_i}-e^{ikr_j})(b_k^\nag-b_{-k}^\dag)}$ and attractive electron-electron interaction
$U_{ij}=- \frac{1}{V} \sum_k e^{-ik (r_j-r_i)} \frac{g_k^2}{\omega_k \sqrt{1-A_k}}$. 

\subsection{Rotating phonon frame}

We construct a unitary transformation to remove the phonon driving following Ref.~\cite{seleznyova_unitary_1995}. We introduce the undriven and driven phonon Hamiltonian, respectively, as
\begin{subequations}
\begin{align}
 H_0 &= P_q P_{-q} + \omega_q^2 Q_q Q_{-q} = \omega_q (b_q^\dag b_q + b_{-q}b_{-q}^\dag) \\ 
 H_1 &= P_q P_{-q} + \omega_q^2 \Omega_t^2 Q_q Q_{-q} = \frac{1}{2}(\Omega_t^2 +1) \omega_q (b_q^\dag b_q + b_{-q}b_{-q}^\dag) + \frac{1}{2}(\Omega_t^2 -1) \omega_q (b_q^\dag b_{-q}^\dag+b_q^\nag b_{-q}^\nag).
\end{align}
\end{subequations}
Applying the unitary transformation, we find
\begin{equation}
 i\frac{d}{dt} W \ket{\psi_0} = i\frac{dW}{dt}  \ket{\psi_0} + W i\frac{d}{dt}  \ket{\psi_0} = i\frac{dW}{dt}  \ket{\psi_0} + W H_0  \ket{\psi_0} = H_1 W \ket{\psi_0}
\end{equation}
yielding
\begin{equation}
 i\frac{dW}{dt}W^\dag + W H_0 W^\dag   = H_1.
 \label{eq:W}
\end{equation}
Multiplying this equation from left by $W$ and from right by $W^\dag$ and identifying $H_0=H_{0,\text{ph}}$ and $H_1=H_{\text{ph,0}} + H_\text{drv}$, we obtain
\begin{equation}
 W^\dag (\hat H_\text{ph,0}+\hat H_\text{drv}) W - i W^\dag \frac{dW}{dt} = \hat H_\text{ph,0}.
 \label{eq:Wt}
\end{equation}

Our goal is to construct a mapping of the quantum problem onto a classical Mathieu equation which determines the transformation $W$~\cite{seleznyova_unitary_1995} uniquely. To this end, we introduce the Ansatz
\begin{equation}
 W(t) = e^{\xi e^{-2i \omega_q t} K_+ - \xi^*  e^{2i \omega_q t} K_-} e^{-2i K_0 \phi} ,
 \label{eq:Wans}
\end{equation}
where 
\begin{align} 
 K_0 = \frac12 (b_q^\dag b_q + b_{-q}b_{-q}^\dag), \qquad K_+ = b_q^\dag b_{-q}^\dag, \qquad  K_- = b_q b_{-q},
\end{align}
which obey SU(2) algebra $[K^-,K^+]=2K_0$, $[K^0,K^{\pm}]=\pm K^{\pm}$. 
The time dependent factors in \eq{eq:Wans} are chosen such that $b_q e^{i \omega_q t}=\hat b_q(t)$ are invariants of the undriven problem $H_0$, which are defined by requiring that they commute with the corresponding action, i.e., $[\hat b_q(t),i \partial_t - H_0]=0$. For a given invariant $\hat b_q(t)$, $\hat b_q(t) \ket{\psi}$ remains an eigenstate of $H_0$ provided $\ket{\psi}$ is an eigenstate, since $0=[\hat b_q(t),i \partial_t - H_0] \ket{\psi} = (i \partial_t - H_0) \hat b_q(t) \ket{\psi}$. We furthermore introduce the invariants of the driven problem as
\begin{equation}
 [\hat a_q(t),i \partial_t - H_1]=0,
 \label{eq:invH1}
\end{equation}
which will be generated by the unitary transformation $\hat a_q(t)=W b_q(t) W^\dag$. By differentiation and subsequent integration we obtain for the invariants
\begin{subequations}
\begin{align}
 \hat a_q(t)=W b_q(t) W^\dag &= e^{i\phi} (\cosh |\xi| b_q e^{i\omega_q t} -\frac{\xi  e^{-i\omega_q t}}{|\xi|} \sinh|\xi| b_{-q}^\dag) \equiv \chi(t) b_q + \lambda^*(t) b_{-q}^\dag \\
 \hat a_{-q}^\dag(t)=W b_{-q}^\dag(t) W^\dag &= e^{-i\phi} (\cosh |\xi| b_{-q}^\dag e^{-i\omega_q t} -\frac{\xi^* e^{i\omega_q t}}{|\xi|} \sinh|\xi| b_{q} ) \equiv \chi^*(t) b_{-q}^\dag + \lambda(t)b_q .
\end{align}
 \label{eq:aTrans}
\end{subequations}
We do not explicitly write the $q$ dependence in $\chi$ and $\lambda$, as these functions are symmetric in $q$, i.e., $\chi_q = \chi_{-q}$.

Next, we compute the invariants $i\partial_t \hat a_q(t)=[H,\hat a_q(t)]$, \eqw{eq:invH1}, using relations \eqw{eq:aTrans}
\begin{subequations}
\begin{align}
 \frac{d \chi}{dt} &=  i \frac{\omega_q}{2} (\Omega_t^2 +1 ) \chi - i \frac{\omega_q}{2} (\Omega_t^2  - 1 ) \lambda^*\\ 
 \frac{d \lambda^*}{dt} &= -i \frac{\omega_q}{2} (\Omega_t^2 +1 ) \lambda^* + i \frac{\omega_q}{2} (\Omega_t^2  - 1 ) \chi.
\end{align} 
\end{subequations}
We transform $\alpha = \chi - \lambda^*$, $\beta = \chi + \lambda^*$ which yields the Mathieu equation
\begin{equation}
 \frac{d^2 \alpha}{dt^2}+\omega_q^2 \Omega_t^2 \alpha=0 \qquad \frac{d\alpha}{dt} = i\omega_q \beta . 
 \label{eq:mathieu}
\end{equation}
From the initial conditions that require $\chi(0)=1$, $\lambda(0)=0$, we find
\begin{equation}
 \alpha(0)=1, \qquad  \dot\alpha(0) = i\omega_q .
\end{equation}
Using the definition of $\chi,\lambda$ we find the relations between $\xi(t)$, $\phi(t)$ from the unitary transformation \eq{eq:Wans} and the parameters of the Mathieu equation
\begin{subequations}
\begin{align}
 \cosh |\xi|  e^{i\phi} &=\frac{e^{-i\omega t}}{2} (\alpha - \frac{i \dot \alpha}{\omega_q}) \\
 \sinh |\xi| \frac{\xi}{|\xi|}   e^{i\phi} &=\frac{e^{i\omega t}}{2} (\alpha + \frac{i \dot \alpha}{\omega_q}).
\end{align}
\end{subequations}

Plugging in the driving from \eq{eq:driveneph}, we find $\Omega_t^2=(1-A_q-A_q \cos 2 \Omega t)$. Even though the Mathieu equation can not be solved analytically, for this form of the driving its solution is well understood, as it realizes the parametric oscillator, which displays a parametric resonance when $\Omega = \omega \sqrt{1-A}$. 
On resonance the phonon squeezing $|\xi|$ increases linearly in time while off resonance it oscillates around a mean value, \fig{fig:mathieu}.

\begin{figure}
        \centering
        \includegraphics[width=0.48\textwidth]{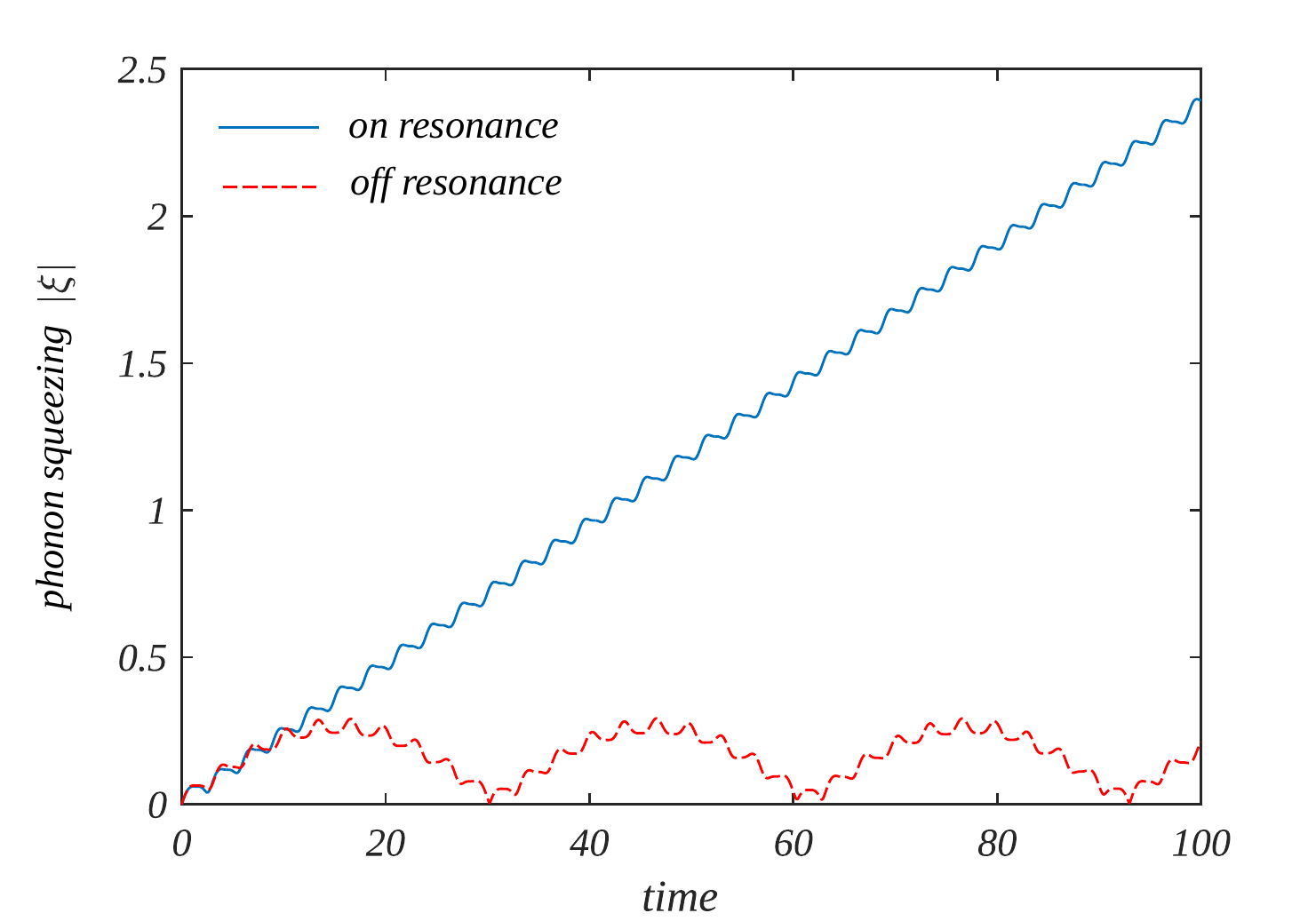}
        \caption{\textbf{Mathieu equation}. Numerical solution of the Mathieu equation at parametric resonance ($A=0.1$, $\Omega=\omega \sqrt{1-A}$), blue solid line, and off resonance ($A=0.1$, $\Omega=0.9 \omega \sqrt{1-A}$), red dashed line. On resonance the phonon squeezing $|\xi|$ increases linearly in time, while off resonance it oscillates around a finite mean value.}
        \label{fig:mathieu} 
\end{figure}

Now we turn again to the electron-phonon problem. Using the unitary transformation $W$, we remove the driving in \eq{eq:driveneph}. Following equation \eq{eq:Wt} we have to transform the phonon operators in the dressed kinetic energy according to $W^\dag\cdot W$
\begin{subequations}
\label{eq:btrans}
\begin{align}
 W^\dag b_q W &=  \cosh |\xi|e^{i\phi} b_q  +\frac{\xi e^{-2i\omega_q t}}{|\xi|} \sinh|\xi| e^{-i\phi} b_{-q}^\dag \\
 W^\dag b_{-q}^\dag W &= \cosh |\xi|e^{-i\phi} b_{-q}^\dag  +\frac{\xi^*e^{2i\omega_q t}}{|\xi|} \sinh|\xi|e^{i\phi} b_{q} . 
\end{align}
\end{subequations}
Assuming, that the drive is adiabatically switched on, the system remains in phonon vacuum in the rotating frame from which we find 
\begin{equation}
 J(t) = - J_0 e^{-\frac{1}{2V} \sum_q (2-2\cos q (r_i -r_j))\frac{g_q^2}{\omega_q^2(1-A_k)^{{3}/{2}}} |\alpha_q|^2},
 \label{eq:timedeppar}
\end{equation}
where $\alpha_q=(\cosh |\xi| - \frac{\xi^*e^{2i\omega_q t}}{|\xi|}\sinh|\xi|)e^{i\phi}$. Upon time averaging this equation yields equation (3) where we parametrize the relative suppression of the electron tunneling compared to the one at equilibrium by the exponential factor $\exp[-\zeta]$.

\begin{figure}
        \centering
        \includegraphics[width=0.98\textwidth]{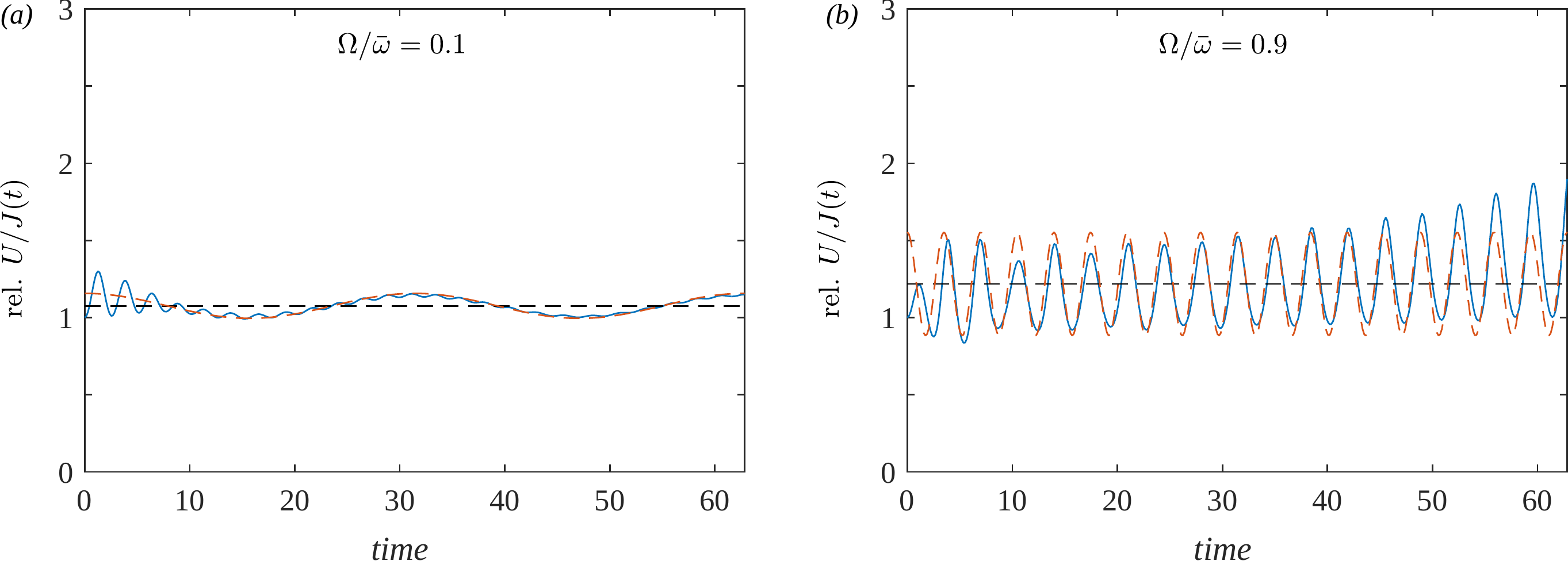}
        \caption{\textbf{Relative strength of the oscillations in the microscopic parameters}. We show the time dependence of $U/J(t)$ relatively to the undriven system for $\bar \omega=U=W/8$, $\Delta \omega=0.2 \bar \omega$, $A=0.1$, \fc{a} off resonance $\Omega/\bar \omega=0.1$ and \fc{b} near the parametric resonance $\Omega/\bar\omega=0.9$ (blue solid line). Initially oscillations at multiple frequencies appear, however, all oscillations except for the driving frequency are washed out in time because for a dispersive phonon the only common frequency of all modes is the driving frequency. Oscillations with $\Omega$ are indicated by the red dashed line which we choose to approximate the dynamics of the microscopic parameters. The time average of $U/J(t)$, black dashed line, is enhanced near resonance and with increasing driving strength.}
        \label{fig:jvt} 
\end{figure}
\subsection{Rescaling of time}

It is convenient to transform the time evolution from the kinetic energy to the interaction energy. Once, we evaluated the phonon dynamics, \eq{eq:timedeppar}, we obtain the effective electron Hamiltonian of the form
\begin{equation}
 H(t) = J(t) \hat H_T - U \hat H_U, 
\end{equation}
where we used $\hat H_T$ and $\hat H_U$ as short hand notation for the kinetic energy and the interaction energy, respectively, and $J(t)$ is the time dependent hopping matrix element. We introduce $ J(t) = J_\text{eq} e^{-\zeta} j(t)$ where $j(t)$ is an oscillating function with mean value one. In order to move the time dependence from the kinetic energy to the interaction energy, we consider
\begin{equation}
 \int_0^t H(t) dt = \int_0^t \tilde H(t) \underbrace{j(t)dt}_{dt'} = \int_0^{t'} \tilde H(f(t')) dt'
 \label{eq:Htrans}
\end{equation}
where 
\begin{equation}
 \tilde H(\tau) = J_\text{eq} e^{-\zeta}  \hat H_T - \frac{U}{j(\tau)} \hat H_U
\end{equation}
and
\begin{equation}
 t'=\int_0^{t'} dt'=\int_0^t j(t)dt = f^{-1}(t).
\end{equation}
The inverse of this equation cannot be calculated analytically, however, for an oscillating function with amplitude small compared to the mean of $j(t)$ follows that $t\sim t'$ which holds because we chose $j(t)$ to oscillate around one. Thus we can directly transform the time dependent part from $J(t)$ to the interaction  and arrive at 
\begin{equation}
 \tilde H(t) =  J_\text{eq} e^{-\zeta} \hat H_T - \frac{U}{j(t)} \hat H_U.
 \label{eq:HElTime}
\end{equation}
As we show in the next section, the oscillations in the interaction can be well described by a single harmonic.

\subsection{Time dependence of the microscopic parameters}

For a dispersive phonon with spread $\Delta \omega$ the time evolution of $U/J(t)$ typically oscillates at driving frequency $\Omega$ which is a common oscillation frequency for the phonon modes at all wave vectors. We approximate the oscillations of $U/J(t)$ with a single harmonic of frequency $\Omega$ and strength $\mathcal{A}$, see \fig{fig:jvt}. In case of strongly off-resonant driving, the interaction oscillates around its equilibrium value, while near the parametric resonance on average it is enhanced, as a result of the polaronic suppression of the bandwidth $J_\text{eq} e^{-\zeta}$. The mean suppressed bandwidth $J_\text{eq} e^{-\zeta}$ and strength $\mathcal{A}$ of the oscillations in $U/J(t)$ are extracted numerically, \textit{cf.} \fig{fig:Uenh}.

\begin{figure}
        \centering
        \includegraphics[width=0.98\textwidth]{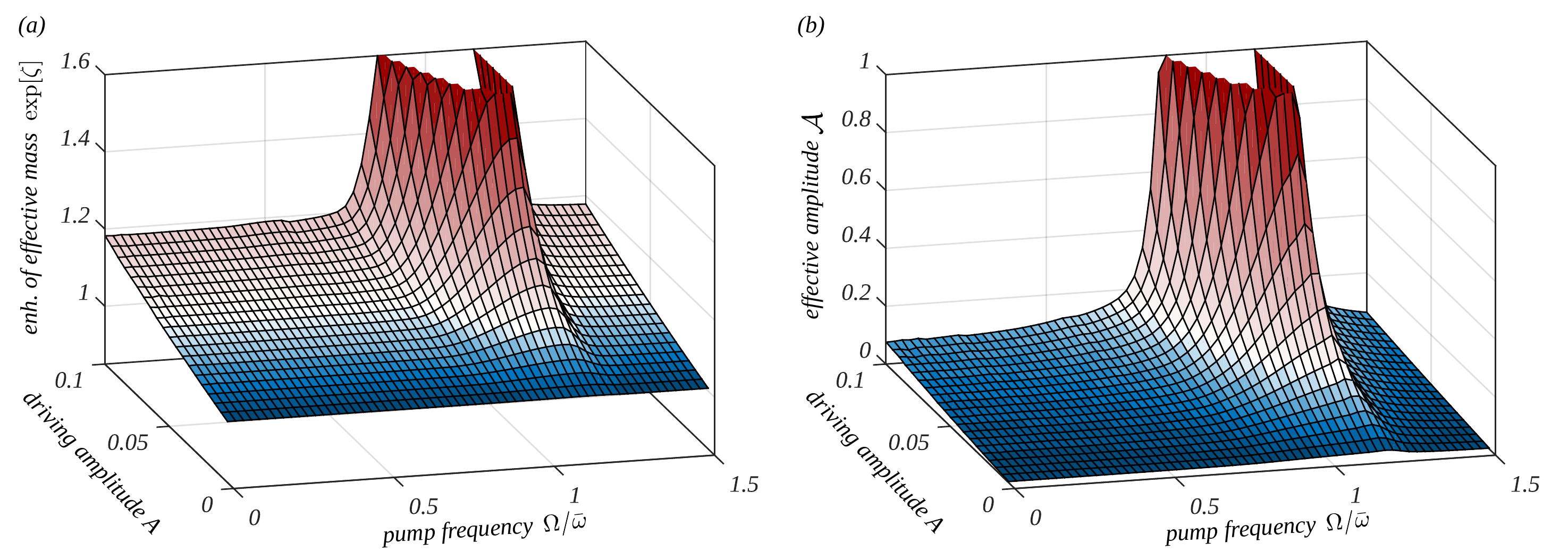}
        \caption{\textbf{Relative enhancement of the effective mass in the driven system compared to the undriven one and the amplitude of coherent oscillations.} The data is shown for linearly dispersing phonons $\bar \omega=U=W/8$, and $\Delta \omega=0.2 \bar \omega$. Near $\Omega/\bar \omega \sim 1$ we find a strong  enhancement of \fc{a} the effective mass $\exp[{\zeta}]$ and \fc{b} the effective amplitude $\mathcal{A}$ due to the efficient phonon squeezing. Furthermore, a weaker enhancement is observed at the higher order resonances where $\Omega/\bar \omega  \sim 1/n$ with $n$ being an integer. In addition to the enhancement of the effective mass near the parametric resonance, it generically increases with increasing driving amplitude $A$ due to the softening of the phonon modes.}
        \label{fig:Uenh} 
\end{figure}

\subsection{Floquet Fermi's Golden Rule \label{app:ffgr}}

Fluctuations around the initial phonon state and the temporal modulation of the interactions \emph{reduce} the quasiparticle lifetime which in turn decrease the superconducting transition temperature. 

Below, we compute the quasiparticle scattering rate from the phonon fluctuations $1/\tau_\text{ph}$ and the modulated interactions $1/\tau_\text{int}$ by Floquet Fermi's Golden Rule. Due to the time dependence of the coupling constants, energy required to create particle-hole excitations can be borrowed from the drive. This enhances the scattering rate compared to equilibrium. 
The total dynamic Cooper pair breaking rate $1/\tau = 1/\tau_\text{ph}+1/\tau_\text{int}$ is shown in \fig{fig:ffgr} for $\bar \omega=U=W/8$. For low driving frequencies, the pair-breaking rate is small, since only higher order Floquet harmonics can provide the required energy and the corresponding matrix elements are small. When the driving frequency is near parametric resonance with phonon pair excitations, the effective interactions and hence the decay rate increases since the drive can efficiently provide the required energy to create particle-hole excitations. In broad regime of parameters we find that the enhanced pair formation rate dominates over the enhanced pair breaking rate, \textit{c.f.} Fig. 1 in the main text.

The full form of the effective Hamiltonian (4) is
\begin{equation}
 \bar H = -J_\text{eq} e^{-\zeta} \sum_{ij\sigma} e^{-\sum_k \alpha_k^*(t)\Gamma_{k}^* b_{k}^\dag} e^{\sum_k \alpha_k(t) \Gamma_k b_{k}^\nag} c_{i\sigma}^\dag c_{j\sigma}^\nag -U (1+\mathcal{A}\cos 2 \Omega t) \sum_i n_{i\uparrow} n_{i\downarrow},
 \label{eq:HphF}
\end{equation}
where $\Gamma_k = -\frac{1}{\sqrt{V}}(e^{ikr_i}-e^{ikr_j}) \frac{g_k}{\omega_k(1-A_k)^{{3}/{4}}}$. Taking the phonon vacuum expectation value of \eq{eq:HphF}, we obtain the first two terms of equation (4). The electron-phonon scattering term, represented by the last term in equation (4) is given by  $\hat H_\text{el-ph scatt.} =-J_\text{eq} e^{-\zeta} \sum_{ij\sigma} (e^{-\sum_k \alpha_k^*(t)\Gamma_{k}^* b_{k}^\dag} e^{\sum_k \alpha_k(t) \Gamma_k b_{k}^\nag} -1) c_{i\sigma}^\dag c_{j\sigma}^\nag$ which as discussed vanishes upon taking the phonon vacuum expectation value. Using a Floquet Fermi's Golden Rule  analysis we estimate the Cooper pair-breaking rate which originates from both (i) phonon fluctuations in the kinetic energy and (ii) modulation of the effective electron-electron interactions.
\begin{figure}
        \centering
        \includegraphics[width=0.48\textwidth]{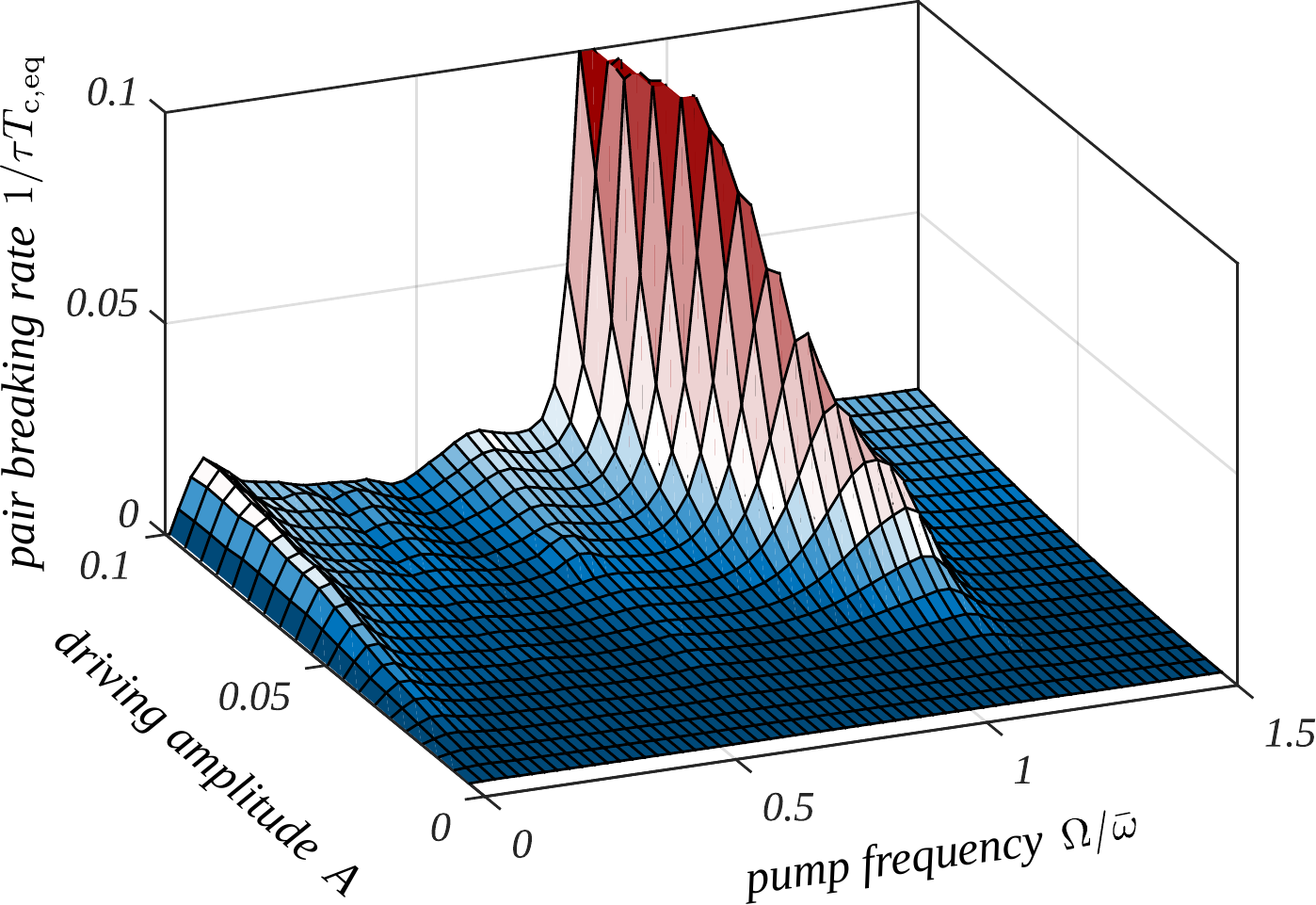}
        \caption{\textbf{Cooper pair breaking rate}. The Cooper pair breaking rate due to phonon fluctuations and the modulation of the effective electron-electron interactions evaluated from Floquet Fermi's golden rule for $\bar \omega=U=W/8$. 
        }
        \label{fig:ffgr} 
\end{figure}

\paragraph{Phonon Fluctuations.}
Fluctuations around the phonon vacuum require energy. This energy can be borrowed from the time dependence of the drive. 
Expanding the exponentials of the electron phonon scattering term $\hat H_\text{el-ph scatt.}$ to first order, we obtain 
\begin{align}
 \hat H_\text{el-ph scatt.} &\sim -J_\text{eq} e^{-\zeta} \frac{1}{\sqrt{V}} \sum_{kq \sigma} \frac{2g_q}{\omega_q(1-A_q)^{{3}/{4}}} [\sum_d\cos k_d - \cos(k_d-q_d)]  (\alpha_q(t) b_q - \alpha_{-q}^*(t) b_{-q}^\dag) c_{k\sigma}^\dag c_{k-q\sigma}^\nag \nonumber \\
 &=\frac{1}{{V}}\sum_{kq \sigma} [\mathcal{F}_{qk} \alpha_q(t) b_{q}^\nag -\mathcal{F}_{qk}^* \alpha_q^*(t) b_{-q}^\dag]\, c_{k\sigma}^\dag c_{k-q\sigma}^\nag
\end{align}
with
\begin{equation}
 \mathcal{F}_{kq}=-{2} J_\text{eq} e^{-\zeta} [\sum_d\cos k_d - \cos(k_d-q_d)] \frac{g_q}{\omega_q(1-A_q)^{{3}/{4}}}  .
\end{equation}
Higher order terms in the series expansion of the exponentials correspond to multi-phonon processes which are energetically suppressed.

The finite lifetime of quasiparticles at the Fermi energy can be calculated from the imaginary part of the retarded phonon-fluctuation self energy  
\begin{equation}
 \frac{1}{\tau_\text{ph}} = \im \Sigma^{\text{ph},+}_{k_F}.
\end{equation}
To leading order the corresponding greater and lesser components are 
\begin{equation}
 \Sigma^{\text{ph},\gtrless}_{k}(t_1,t_2) = \eqfigscl{0.75}{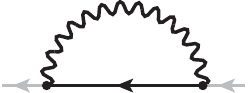} = \frac{i}{{V}} \sum_q |\mathcal{F}_{kq}|^2 \alpha_q(t_1) \alpha_q^*(t_2) G_{k-q}^\gtrless(t_1,t_2) D_q^\gtrless (t_1,t_2).
 \label{eq:seph}
\end{equation}
Here, $G_k^\gtrless(t_1,t_2)$ is the bare electron and $D_q^\gtrless(t_1,t_2)$ the bare phonon propagators, which are functions of the time difference as the unperturbed Hamiltonian is static. Thus, in Fourier space they are given by 
\begin{align*}
 G^>_k(\omega) &\approx -2\pi i \delta(\omega-\epsilon_k+\mu) [1-n_k], \\
 G^<_k(\omega) &\approx +2\pi i \delta(\omega-\epsilon_k+\mu) n_k, \\
 D^>_q(\omega) &\approx -{i\pi} \delta(\omega-\omega_q), \\
 D^<_q(\omega) &\approx -{i\pi} \delta(\omega+\omega_q),
\end{align*}
where we considered the phonons to be in the vacuum state and thus neglected their distribution functions in $D^>$ and $D^<$. The self-energy contains the drive $\alpha_q(t)$ and therefore is a fully non-equilibrium object which is a function of two times, \eq{eq:seph}. We rewrite the self-energy using the average time $T=\frac{1}{2}(t_1+t_2)$ and the time difference $t=t_1-t_2$
\begin{equation}
 \Sigma^{\text{ph},\gtrless}_{k}(t,T)= \frac{i}{{V}} \sum_q |\mathcal{F}_{kq}|^2 \alpha_q(T+t/2) \alpha_q^*(T-t/2) G_{k-q}^\gtrless(t) D_q^\gtrless (t).
\end{equation}
In order to estimate a quasi-particle lifetime, we integrate over the ``slow'' timescale $T$ which yields for the effective coupling $|\bar{\alpha}_q(t)|^2 = \frac{\Omega}{\pi} \int_0^{\frac{\pi}{\Omega}}dT \alpha_q(T+t/2)\alpha_q^*(T-t/2)$. The effective coupling $|\bar{\alpha}_q(t)|^2$, distinguished from the bare coupling by the bar, is only a function of the time difference $t$. From a Fourier transform we obtain its Floquet components $|\bar{\alpha}_{qn}|^2$, where $n$ ranges from $-\infty$ to $\infty$. 
We calculate the retarded self-energy $\Sigma^+_{\text{ph},k}$ using 
\begin{equation}
 \Sigma^{\text{ph},+}_{k}(\omega) = i \int \frac{d \omega'}{2\pi} \frac{\Sigma^{\text{ph},>}_{k}(\omega')-\Sigma^{\text{ph},<}_{k}(\omega')}{\omega-\omega'+i0^+},
\end{equation}
where $\omega$ is the conjugate variable to the time difference $t$, and find for the lifetime at the Fermi surface
\begin{equation}
 \frac{1}{\tau_\text{ph}} =  \frac{\pi}{2V}\sum_{qn}  |{\mathcal{F}}_{qk_F}|^2 |\bar \alpha_{qn}|^2 \{ (1-n_{k_F-q}) \delta(2 n \Omega - E_{k_F-q}-\omega)+n_{k_F-q} \delta(2 n \Omega - E_{k_F-q}+\omega)\}.
\end{equation} 
In order to obtain a semi-analytical estimate for the pair breaking rate, we neglect the weak wavevector dependence of $g_q/\omega_q(1-A_q)^{{3}/{4}}$ and replace them by their mean. We replace wavevector summations by integrals over energies with a constant density of states, yielding
\begin{align}
 \frac{1}{\tau_\text{ph}} =  \frac{\pi}{2}\sum_{n>0} \frac{e^{\zeta}}{8J_\text{eq}} &\frac{\bar g^2}{\bar\omega^2(1-A)^{{3}/{2}}} (2n\Omega-\bar\omega)^2 \Theta(2n\Omega - \bar\omega) \{ \alpha_n \Theta(|E_F+2n\Omega-\bar\omega|-W)\nonumber\\
 &+\alpha_{-n} \Theta(|E_F-2n\Omega+\bar\omega|-W)\},
 \label{eq:ffgrapp}
\end{align}
where $W=4J_\text{eq} e^{-\zeta}$ is half of the electronic bandwidth.

\paragraph{Modulated interactions.}
The temporal modulation of the effective electron-electron interaction leads to another source for decreasing the quasiparticle lifetime. Similarly as in the case of phonon fluctuations, we estimate the interaction decay rate by computing the imaginary part of the leading order self-energy contribution 
\begin{align}
 \Sigma^{\text{int},\gtrless}_{k}(t_1,t_2) = \eqfigscl{0.75}{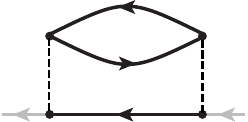} = \frac{1}{{V^2}} \sum_{ql} &U^2(1+\mathcal{A}\cos 2\Omega t_1) (1+\mathcal{A}\cos 2\Omega t_2) \times \nonumber\\ & G_{k+l-q}^\gtrless(t_1,t_2) G_{q}^\gtrless(t_1,t_2) G_l^\lessgtr (t_2,t_1).
 \label{eq:seit}
\end{align}
Performing a Floquet Fermi's Golden Rule analysis, we first integrate over the ``slow'' timescale $T$ to obtain an effective coupling as a function of the time difference $t$ and then compute the Floquet components of the coupling. Plugging this into the expression for the retarded self-energy and taking the imaginary part, we obtain the decay rate
\begin{align}
 \frac{1}{\tau_\text{int}} = \frac{U^2 \mathcal{A}^2}{4 V^2} &\sum_{ql} [(1-n_{k_F+l-q})\, (1-n_q) \,n_l + n_{k_F+l-q}\,n_q \,(1-n_l) ] \nonumber \\ & [\delta(2\Omega - (E_{k_F+q-l} +E_{q}-E_l)+\delta(2\Omega + (E_{k_F+q-l} +E_{q}-E_l)].
\end{align}

We evaluate both quasiparticle decay rates numerically, \textit{cf.} \fig{fig:ffgr}. 
The total pair-breaking rate $1/\tau = {1}/{\tau_\text{ph}}+{1}/{\tau_\text{int}}$, which we consider as an imaginary self-energy correction in the Floquet BCS equations, is in a wide parameter range much smaller than the Cooper pair formation rate and thus only slightly shifts the transition temperature to lower values.

\section{Floquet BCS approach \label{app:bcs}}

We first evaluate the equations of motion for $c_{k\uparrow}^\dag c_{-k\downarrow}^\dag$ from the rescaled Hamiltonian equation (4) taking into account the Cooper pair breaking rate $1/\tau$, computed in Sec.~\ref{app:ffgr}, as an imaginary self-energy correction
\begin{equation}
 \frac{d}{dt} c_{k\uparrow}^\dag c_{-k\downarrow}^\dag = 2i (\epsilon_k+i/\tau-\mu) c_{k\uparrow}^\dag c_{-k\downarrow}^\dag - i \frac{U(1+\mathcal{A}\cos 2 \Omega t)}{V} \sum_{mq} c_{m\uparrow}^\dag c_{q\downarrow}^\dag (c_{m+q-k \downarrow}c_{-k \downarrow}^\dag-c_{k \uparrow}^\dag c_{m+q+k \uparrow})
\end{equation}
and factorize the quartic term using a mean-field decoupling 
\begin{align}
  \frac{d}{dt}\langle  c_{k\uparrow}^\dag c_{-k\downarrow}^\dag \rangle &=  2i(\epsilon_k+i/\tau-\mu) \langle  c_{k\uparrow}^\dag c_{-k\downarrow}^\dag \rangle \nonumber\\ &- i \frac{U(1+\mathcal{A}\cos 2 \Omega t)}{V} \sum_{q} \langle c_{k\uparrow}^\dag c_{-k\downarrow}^\dag \rangle (n_{q\uparrow} + n_{q\downarrow}) + \langle c_{q\uparrow}^\dag c_{-q\downarrow}^\dag \rangle (1-n_{k\uparrow}-n_{k\downarrow}).
  \label{eq:app:eomMF}
\end{align}
Using spin symmetry and defining $2\rho=\frac{1}{V}\sum_q (n_{q\uparrow} + n_{q\downarrow})$ we obtain
\begin{equation}
  \frac{d}{dt} \langle  c_{k\uparrow}^\dag c_{-k\downarrow}^\dag \rangle = 2i (\epsilon_k+i/\tau-\mu-U(1+\mathcal{A}\cos 2 \Omega t)\rho) \langle  c_{k\uparrow}^\dag c_{-k\downarrow}^\dag \rangle - i U(1+\mathcal{A}\cos 2 \Omega t) (1-2n_{k}) \frac{1}{V}\sum_{q} \langle c_{q\uparrow}^\dag c_{-q\downarrow}^\dag \rangle.
\end{equation}
Next, we remove the term $2i U\rho\mathcal{A}\cos 2 \Omega t  \langle c_k^\dag c_{-k}^\dag \rangle$ by an appropriate unitary transformation of the form
\begin{equation}
 \langle c_{k\uparrow}^\dag c_{-k\downarrow}^\dag \rangle = a_k^* \exp[-{i \frac{U\rho\mathcal{A}}{\Omega}\sin 2 \Omega t}]
\end{equation}
which gives:
\begin{equation}
 \frac{d}{dt} a_k^* = 2i (\epsilon_k+i/\tau-\mu-U \rho) a_k^* - iU (1+\cos2\Omega t)  (1-2n_{k})\frac{1}{V}\sum_{q} a_q^*.
\end{equation}
Using the Floquet Ansatz
\begin{equation}
 a_k^*(t) = e^{i E t} \sum_n a_{kn}^* e^{i2n\Omega t}
\end{equation}
we obtain
\begin{equation}
 [E+2 n\Omega - 2 (\epsilon_k+i/\tau -\mu - U \rho) ]a_{kn}^*  = -(1-2n_k)\frac{U}{V} \left[ \sum_q a_{qn}^* + \frac{\mathcal{A}}{2}\sum_q (a_{qn+1}^* + a_{qn-1}^*)\right].
 \label{eq:app:floquet}
\end{equation}
Dividing by $E+2 n\Omega - 2 (\epsilon_k+i/\tau -\mu - U \rho)$, and summing over $k$, we find the Floquet BCS gap equation:
\begin{equation}
 \frac{1}{V}\sum_k a_{kn}^* = -\underbrace{\frac{1}{V}\sum_k \frac{1-2n_k}{E+2 n\Omega - 2 (\epsilon_k+i/\tau -\mu - U \rho) }}_{=F_n}\frac{U}{V}\left[ \sum_q a_{qn}^* + \frac{\mathcal{A}}{2} \sum_q (a_{qn+1}^* + a_{qn-1}^*)\right].
\end{equation}
Defining the gap $\Delta_n=\frac{U}{V}\sum_k a_{kn}^*$, we obtain the simple system of equations
\begin{equation}
(U^{-1}+F_n)\Delta_n + \frac{\mathcal{A}}{2}F_n(\Delta_{n-1}+\Delta_{n+1}) =0 .
\label{eq:app:gap}
\end{equation}
The fact that we used a single harmonic to describe the time evolution of $U(t)$ reflects in the gap equation having only a single side band. More complicated functions would lead to further side bands which would give quantitative differences but our conclusions will not be altered on the qualitative level.
The BCS Floquet equations have a nontrivial solution, when the determinant is zero, which we determine by scanning $E$ in the complex plane. The Cooper pair formation rate is characterized by the imaginary part of $E$.

\end{onecolumngrid}

%

\end{document}